\documentstyle[aps]{revtex}
\input{epsf.tex}
\begin{document}
\draft
\title{Transition to the Fulde-Ferrel-Larkin-Ovchinnikov planar phase : 
a quasiclassical investigation with Fourier expansion}
\author{R. Combescot and C. Mora}
\address{Laboratoire de Physique Statistique,
 Ecole Normale Sup\'erieure*,
24 rue Lhomond, 75231 Paris Cedex 05, France}
\date{Received \today}
\maketitle

\begin{abstract}
We explore, in three spatial dimensions, the transition from the normal state to the Fulde-Ferrel-Larkin-Ovchinnikov superfluid phases. We restrict ourselves to the case of the 'planar' phase, where the order parameter depends only on a single spatial coordinate. We first show that, in the case of the simple Fulde-Ferrell phase, singularities occur at zero temperature in the free energy which prevents, at low temperature, a reliable use of an expansion in powers of the order parameter. We then introduce in the quasiclassical equations a Fourier expansion for the order parameter and the Green's functions, and we show that it converges quite rapidly to the exact solution. We finally implement numerically this method and find results in excellent agreement with the earlier work of Matsuo \emph{et al}. In particular when the temperature is lowered from the tricritical point, the transition switches from first to second order. In the case of the first order transition, the spatial dependence of the order parameter at the transition is found to be always very nearly a pure cosine, although the maximum of its modulus may be comparable to the one of the uniform BCS phase.
\end{abstract}


\section{INTRODUCTION}

Despite being actively investigated for fourty years the problem of the
structure of the superconducting order parameter in very high fields is still
the subject of intensive research. In the compounds of interest the
coupling of the magnetic field to electronic spins can no longer be ignored,
and the situation where it is the only relevant one has to be considered.
In this case one faces the problem of pairing electrons, for which the
spin up and spin down chemical potentials are not the same. This question 
has been addressed independently by Fulde and Ferrell \cite{ff} (FF) and 
by Larkin and Ovchinnikov \cite{larkov} (LO), who proposed that the best
order parameter corresponds to pairs formed with a non zero total momentum,
in contrast to the standard situation of the BCS theory. It is worth
noting that this kind of problem has been found recently to be quite relevant
for ultracold atomic Fermi gases \cite{rcff} as well as for the physics
of neutron stars \cite{bowers,wilcz,canar}. More specifically Larkin and
Ovchinnikov \cite{larkov} considered for the order parameter superpositions  
of different plane waves, corresponding physically to different pair 
total momentum. They investigated which superposition was favored near
the transition at $T=0$. Nevertheless they considered only a second
order phase transition, which is not the most general situation as we will
discuss below. Hence 
the question of the exact structure of the order parameter in the FFLO
phases is still an open problem. Since most experiments identifying
tentatively FFLO phases rest heavily on the theoretical analysis, this
is also a problem with major experimental implications.

In a preceding paper \cite{cm}  we have investigated analytically the 
transition to the FFLO phases in the vicinity of the tricritical point 
(TCP), where the FFLO transition line starts. This point is located at 
$T_{\rm{tcp}} / T_{c0} = 0.561 $ where $T_{c0}$ is the critical 
temperature for  $ \bar{ \mu } = 0 $ , with $ 2 \bar{ \mu } = \mu 
_{\uparrow} - \mu_{\downarrow} $ being the chemical potential 
difference between the two fermionic populations forming pairs, as for
example spin up and down electrons (the 
corresponding effective field is  $ \bar{ \mu }_{\rm{tcp}}/ T_{c0} = 
1.073 $). In agreement with preceding numerical work 
\cite{buz1,buz2,matsuo} we have found that the transition is first order 
to an order parameter which is, to a very good precision, simply 
proportional to a one-dimensional 'planar' texture $\Delta ( {\bf  r}) 
\sim \cos({\bf  q}. {\bf  r})$. This order parameter is actually the one 
which has been shown by Larkin and Ovchinnikov \cite{larkov} 
to be the most favorable for a (second order) transition at $T = 0$. 
Compared with other works we have been able to understand 
qualitatively and quantitatively the reasons which favor this order 
parameter with respect to all the other possible ones. Namely we have
shown that a real order parameter is favored, and that, among these states, 
those with the smallest number of plane waves are prefered. This then leads 
to an order parameter with a $\cos({\bf  q}_{0}. {\bf  r})$ dependence, in
agreement with preceding work.

A remarkable feature of the results is that the location in the $ \bar{ \mu 
},T$ plane of this first order transition toward the 'planar' order
parameter is very near the standard FFLO 
second order phase transition. This is true not only near the TCP 
\cite{cm,buz1,buz2} but also almost down to $T = 0$ \cite{matsuo} 
(actually the transition to the 'planar' order
parameter goes back to a second order phase transition at 
low temperature in agreement with  LO). This proximity of a second 
order transition may lead to believe that the order parameter $ \Delta $ is 
reasonably small at the first order transition. This is trivially valid near 
the TCP where a Landau-Ginzburg type expansion up to sixth order in 
order parameter $ \Delta $ could be performed \cite{cm,buz1}, but it is 
a tempting hypothesis even at lower temperature. This possibility has 
been explored by Houzet et al. \cite{buz2}. They found problems in 
applying this scheme because, for the stablest phase, namely the planar 
one, the coefficient of the sixth order in $ \Delta $ changes sign 
when the temperature is lowered not much below the TCP. This leads 
to an instability and so the expansion up to sixth order in powers of $ 
\Delta $ becomes inconsistent.

We will first analyze this problem and show that it is already present in 
the simple case of the Fulde-Ferrell phase where the order parameter is 
given by $ \Delta ( {\bf  r}) = \Delta   \exp(i{\bf  q}.{\bf  r})$. This 
analysis gives a clear hint that an expansion in powers of $ \Delta $ is 
going to fail anyway at low temperature. This suggest that one should 
avoid to perform such an expansion. Among the various possibilities 
for improving the situation, one of them is to remark that, at the 
transition, near the TCP, the actual order parameter is quite near a 
simple superposition of plane waves \cite{cm} even if the transition is 
first order (for a second order transition the order parameter is exactly 
such a superposition, as investigated \cite{larkov} for example by 
Larkin and Ovchinnikov at $T=0$). So the power expansion near the 
TCP amounts also to keep only the lowest order in a Fourier expansion 
of the order parameter. This leads to look for a Fourier expansion in the 
equations instead of a $ \Delta $ expansion.

Since we want to deal with the full nonlinear, space-dependent 
problem, the convenient starting point is not Gorkov's equations, but 
rather the quasiclassical equations of Eilenberger \cite{eil1}, Larkin and 
Ovchinnikov \cite{larkov1}. Not only are those equations in their 
simplest form the most compact and convenient formulation of our 
problem, but a major advantage is that they can be extended in full 
generality to much more complex situations \cite{sererain} and allow to 
formulate transport problems, including many-body effects, with the 
same level of efficiency. However since we have to deal with a 
comparatively simpler problem, we will use for simplicity in this paper 
the original formulation and notations of Eilenberger \cite{eil1}. In 
comparison the general formalism is used by Burkardt and Rainer 
\cite{br} for an analysis of a FFLO transition in two dimensions with a 
planar order parameter. 

In this paper we will show that the introduction of a Fourier expansion 
in the quasiclassical equations allows to obtain a solution which 
converges very rapidly toward the exact result. As a consequence a few 
terms in the expansion provide an excellent approximation. Here we 
will just deal with the principle of this method and its application to the 
planar transition. In particular we will rederive the results of Matsuo et 
al \cite{matsuo}. Applications to other more complex cases, which are the more fundamental interest of this procedure, will be considered in another paper.

The paper is organized as follows. In the next section we consider the free energy and study in particular  the simple case of the Fulde-Ferrell phase, and show that it has singularities at $T = 0$ which make an expansion in powers of the order parameter unreliable. In section \ref{fourier} we explain our Fourier expansion for the simplest case of a cosine order parameter. This is then generalized in the following section to any (one-dimensional) order parameter. Finally we give in section \ref{num} the results of the numerical implementation of our method.

\section{THE  FULDE-FERRELL  PHASE  T = 0  FREE  ENERGY}

We will show that the problems arising in the expansion of the free 
energy in powers of the order parameter are already present when one 
considers the simple Fulde-Ferrell (FF) state. Let us start with the 
completely general expression, that we will use further on, for the free 
energy difference per unit volume between the superconducting state 
and the normal state \cite{br,eil,werth} :
\begin{eqnarray}
\Omega _{s} - \Omega _{n}=  \int d{\bf  r}  \frac{1}{V} |  \Delta ( {\bf  
r}) | ^{2}
+ 4 \pi  T N _{0} {\mathrm{Re}} \sum_{n=0}^{ \infty} 
\int_{\bar{\omega  }_{n}}^{ \infty}
d \omega \int \frac{d \Omega _{k}}{4 \pi } [ g _{s} ( \omega , \hat{ 
{\bf  k} }, {\bf  r})
- g _{n} ( \omega , \hat{ {\bf  k} }, {\bf  r})] 
\label{eq1}
\end{eqnarray}
Here $ V$ is the standard BCS interaction and $ N _{0}$ is the single 
spin density of states at the Fermi energy. The difference $ \mu 
_{\uparrow} - \mu_{\downarrow} = 2 \bar{ \mu }  $ between spin up 
and spin down chemical potentials comes in the definition of $ 
\bar{\omega  }_{n} = \omega _{n} - i \bar{\mu }$ where $ \omega 
_{n} = \pi T (2n+1) $ are Matsubara frequencies. For the " $ \xi $-
integrated " or quasiclassical Green's functions we have used 
Eilenberger's notations $ g ( \omega , \hat{ {\bf  k} }, {\bf  r}) =  
\frac{i}{ \pi } \int d \xi _{k} G( \omega , {\bf  k}, {\bf  r}) $ where 
$\xi _{ {\bf  k}}$ is the kinetic energy measured from the average 
Fermi level $ (1/2) ( \mu _{\uparrow}+ \mu_{\downarrow})$ and $ 
G(\omega _{n} , {\bf  k}, {\bf  r}) $ is the usual temperature Green's 
function (these Green's functions we deal with are those for up spin, 
the down spin Green's functions are obtained by a simple transform 
and the sum over the spin leads to take the real part in Eq.(1)). With 
these notations we have $ g _{n} (\omega _{n} , \hat{ {\bf  k} }, {\bf  
r}) = 1$ for $ \omega _{n} > 0 $. It results directly from the starting 
GorkovÕs equations \cite{larkov} that the Green's functions in the 
presence of the effective field $ \bar{\mu }$ are obtained from those in 
the absence of it by the simple replacement of $ \omega _{n}$ by $ 
\bar{\omega  }_{n} $.

We look now for the expression of this free energy at $ T = 0 $ for the 
simple FF state where the order parameter is given by $ \Delta ( {\bf  
r}) = \Delta   \exp(i{\bf  q}.{\bf  r})$. Since in this case $  |  \Delta ( 
{\bf  r}) | ^{2} =  \Delta ^{2}$, the Green's function is just obtained 
from the standard BCS one by shifting \cite{larkov} all the momenta by 
$ {\bf  q}/2$. Finally the quasiclassical Green's function is just the 
BCS one, except that we have to change $ \omega $ into $ \omega  - i 
\bar{\mu } _{k}$ with $ \bar{\mu } _{k} = \bar{\mu } ( 1 -  \bar{q } \, 
\hat{ {\bf  k} }.\hat{ {\bf  q} })$, where we have defined the reduced 
wavevector $\bar{q }= q k _{F}/(2m \bar{\mu })$ (this results also 
from the general Eilenberger's equations we will write below).

The free energy for the standard uniform BCS phase with $ \bar{\mu 
}= 0 $ is :
\begin{eqnarray}
\Omega \equiv  \Omega _{s} - \Omega _{n}= \frac{1}{2} N _{0} 
\Delta _{0}^{2}
x ^{2} \ln (x ^{2}/e)
\label{eq2}
\end{eqnarray}
where $ \Delta _{0} = 2 \omega _{c} \exp(-1/ N _{0}V)$ is the zero 
temperature BCS phase gap ($ \omega _{c}$ is the standard cut-off of 
BCS theory), and we have expressed $ \Delta $ in units of $ \Delta 
_{0}$ by introducing $ x =  \Delta / \Delta _{0}$. This free energy is 
naturally minimum for $x=1$ and the minimum is $- \frac{1}{2} N 
_{0} \Delta _{0}^{2}$. In the presence of a nonzero effective field $ 
\bar{\mu }>0$ this expression becomes from Eq.(1): 
\begin{eqnarray}
\frac{\Omega}{ N _{0} \Delta _{0}^{2}} = - \frac{ x ^{2}}{2} +  
\bar{m}^{2}
+ {\mathrm{Re}} [x ^{2} \ln (\bar{m}+\sqrt{\bar{m}^{2}- x ^{2}})-
\bar{m} \sqrt{\bar{m}^{2}- x ^{2}}]
\label{eq3}
\end{eqnarray}
where we have also expressed $ \bar{\mu }$ in units of $ \Delta _{0}$ by 
setting $  \bar{m} = \bar{\mu } / \Delta _{0}$. For $ \Delta > \bar{\mu 
}$ this free energy reduces to $ \frac{\Omega}{ N _{0} \Delta 
_{0}^{2}}= \frac{1}{2} x ^{2} \ln (x ^{2}/e) +  \bar{m}^{2}$ and 
gives the standard Clogston-Chandrasekhar \cite{clog,chand} first 
order transition $ \bar{m} = 1/ \sqrt{2}$. On the other hand it gives for 
small $ \Delta \ll \bar{\mu }$ the expansion $ \frac{\Omega}{ N _{0} 
\Delta _{0}^{2}}=  x ^{2} \ln (2  \bar{m}) - x ^{4}/8 \bar{m}^{2} - 
x ^{6}/32 \bar{m}^{4}$, leading in particular to the second order 
spinodal transition for $ \bar{m} = 1/2$. This expansion can be 
generalized at $ T \neq 0$ as :
\begin{eqnarray}
\frac{\Omega}{ N _{0}} =   \ln [T/ T _{sp}(\bar{\mu }/T)]  \: \Delta  
^{2} + \sum_{p=1}^{ \infty} 
(-1) ^{p+1} \frac{(2p)!}{2 ^{2p} p! (p+1)!} A _{2p} \Delta  ^{2p+2}
\label{eq4}
\end{eqnarray}
with :
\begin{eqnarray}
A _{2p} =  2 \pi T \: {\mathrm{Re}}
 [ \sum_{n=0} ^{ \infty} \frac{1}{\bar{\omega  }_{n} ^{2p+1}} ]
\label{eq5}
\end{eqnarray}
and $ T _{sp}(\bar{\mu }/T)$ is the temperature of the second order 
spinodal transition toward the standard BCS phase. It is interesting to 
note that, while the coefficients $ A _{2p}$ are clearly all
positive when $ \bar{\mu } \rightarrow 0 $, they are given by $ A 
_{2p} = (-1) ^{p}/(2p \bar{\mu } ^{2p})$ when $ T \rightarrow 0 $. 
Moreover  one can see that $ A _{2p}$ has $p$ zeros when $ 
\bar{\mu }/T$ goes from $0$ to $ \infty$ (one goes basically from $ A 
_{2p}$ to $ A _{2p+2}$ by taking a double derivative with respect to 
$\bar{\mu }$). Hence the higher order coefficients have many changes 
of sign in the low temperature range. This feature corresponds to the 
singular behaviour which occurs for $ \Delta = \bar{\mu }$ at $ T = 0 $ 
in Eq.(3). It allows also to understand that the change of signs found by 
Houzet et al. \cite{buz2} are not simple accidents, but a systematic 
behaviour linked to the singularity appearing at  $ T = 0 $.

Finally the $ T = 0 $ free energy of the FF phase is obtained by 
replacing in Eq.(3) $ \bar{\mu }$ by $ \bar{\mu } _{k}= \bar{\mu } ( 1 
-  \bar{q } \,  \hat{ {\bf  k} }.\hat{ {\bf  q} })$ and averaging over the 
direction $ \hat{ {\bf  k} }$ as in Eq.(1). We give the result only in the 
case where $\bar{q } > 1$ since this is the range of wavevector 
corresponding to the actual minimum of the free energy. One finds :
\begin{eqnarray}
\frac{\Omega}{ N _{0} \Delta _{0}^{2}} = - \frac{ x ^{2}}{2} +  
\bar{m}^{2} (1+ \frac{\bar{q } ^{2}}{3} )
+  \frac{ x ^{2}}{2 \bar{m} \bar{q }}{\mathrm{Re}} [\bar{m} _{+} 
\ln (\bar{m} _{+}+\sqrt{\bar{m} _{+}^{2}- x ^{2}}) - \sqrt{\bar{m} 
_{+}^{2}- x ^{2}} + (\bar{m} _{+} \rightarrow \bar{m} _{-})]   
\nonumber  \\
-  \frac{ 1}{6 \bar{m} \bar{q }} {\mathrm{Re}} [ (\bar{m} 
_{+}^{2}- x ^{2}) ^{3/2} + (\bar{m} _{+} \rightarrow \bar{m} _{-
})]
\label{eq6}
\end{eqnarray}
where we have used the notation $ \bar{m} _{ \pm } \equiv \bar{m} 
(\bar{q } \pm 1)$. 

This result has singularities for $  \Delta = \bar{\mu }(\bar{q } \pm 1)$. 
These are just the manifestation of the singularity found in Eq.(3) for $  
\Delta = \bar{\mu }$, corresponding to the upper and lower bound in 
the angular integration. In particular the singularity at $  \Delta = 
\bar{\mu }(\bar{q } - 1)$ gives the radius of convergence of the 
expansion in powers of $ \Delta $. A particular consequence is that no expansion is 
possible for $\bar{q } = 1$. This is just the situation which is found when
one works in a two-dimensional space. This singular situation leads to a cascade
involving an infinite number of phase transitions when the temperature goes
to zero, as we have shown elsewhere \cite{fflo2d}. In the case of a three-dimensional
space, with which we deal in this paper, the radius of convergence is non zero,
but it is fairly small since the minimum free energy is found at low temperature 
for values of $\bar{q }$ not far from the $T = 0$ LO result $\bar{q }= 
1.2 $. Therefore a rapidly convergent expansion for the free energy is 
only valid for quite small $ \Delta $, and this happens to be in 
contradiction with the values of $ \Delta $ needed to minimize this free 
energy. Naturally this expansion of Eq.(6) can be performed explicitely 
and the problem with the convergence is then quite obvious.

Now it is clear that these same problems arise if, instead of a phase with 
a single plane-wave as is the FF phase, we consider a more complicated 
phase which is a sum of plane waves, such as the planar phase $\Delta ( 
{\bf  r}) \sim \cos({\bf  q}. {\bf  r})$. This is already obvious from the 
fact that the terms which arise in the expansion for the FF phase will 
also appear in the expansion for this phase. Other terms with weaker 
singularities at $  \Delta = \bar{\mu }(\bar{q } - 1)$ will also be 
present. We note that a singularity is already present in the fourth order 
terms investigated \cite{larkov} by LO, as it can be seen from the 
explicit expression of their integral J, but it occurs for a specific value 
of the angle between the wavevectors which happens to be irrelevant for 
their final conclusion. Therefore we come to the conclusion that, due to 
the singular behaviour which occurs at $ T = 0 $, we can not rely 
anymore on an expansion in powers of $ \Delta $ when the temperature 
is lowered. It is conceivable that such an expansion could still be proper 
by accident for a specific phase, but it is unsafe for a general 
exploration of the various phases in competition. A possible partial cure 
for this problem could be to sum up the most divergent contributions, 
which are precisely those occuring in the FF phase. We have tried such 
an approach, but although it provides some improvement it clearly does 
not lead to a satisfactory situation.

Therefore in an attempt to extend the simple approach around the TCP 
we will in the next section proceed to a Fourier expansion in the exact 
quasiclassical formulation of the problem. This will prove to be 
completely satisfactory.

\section{FOURIER  EXPANSION}\label{fourier}

We start from Eilenberger's equations for the diagonal $ g ( \omega , 
\hat{ {\bf  k} }, {\bf  r})$ and off-diagonal $ f ( \omega , \hat{ {\bf  k} 
}, {\bf  r})$ quasiclassical propagators, which we simplify from the 
outset by taking $ \hbar = 1$ and $ m = 1/2 $. They read \cite{eil1} :
\begin{eqnarray}
( \omega + {\bf  k}.{\bf \nabla}) f ( \omega , \hat{ {\bf  k} }, {\bf  r}) 
= \Delta ( {\bf  r}) 
g ( \omega , \hat{ {\bf  k} }, {\bf  r}) \nonumber  \\
( \omega - {\bf  k}.{\bf \nabla}) f ^{+} ( \omega , \hat{ {\bf  k} }, {\bf  
r}) = \Delta ^{*}( {\bf  r}) 
g ( \omega , \hat{ {\bf  k} }, {\bf  r}) \nonumber  \\
2{\bf  k}.{\bf \nabla} g ( \omega , \hat{ {\bf  k} }, {\bf  r}) = \Delta 
^{*}( {\bf  r}) 
f ( \omega , \hat{ {\bf  k} }, {\bf  r}) -  \Delta ( {\bf  r}) 
f ^{+} ( \omega , \hat{ {\bf  k} }, {\bf  r})
\label{eq7}
\end{eqnarray}
where $ {\bf k }$ is at the Fermi surface $ k = k_F$. 
Actually $g$ is given in terms of $f$ and $f ^{+}$ by the normalization 
condition :
\begin{eqnarray}
g ( \omega , \hat{ {\bf  k} }, {\bf  r}) = (1 - f ( \omega , \hat{ {\bf  k} 
}, {\bf  r}) 
f ^{+} ( \omega , \hat{ {\bf  k} }, {\bf  r}))^{1/2}
\label{eq8}
\end{eqnarray}
so the last equation results from the two first ones. These ones are also 
related \cite{eil1} since :
\begin{eqnarray}
f  ^{*} (- \omega , \hat{ {\bf  k} }, {\bf  r}) =
f ^{+} ( \omega , \hat{ {\bf  k} }, {\bf  r})  && \hspace{3cm}
g ^{*} (- \omega , \hat{ {\bf  k} }, {\bf  r}) =
- g ( \omega , \hat{ {\bf  k} }, {\bf  r})  \nonumber  \\
f  ^{*} (\omega , - \hat{ {\bf  k} }, {\bf  r}) =
f ^{+} ( \omega , \hat{ {\bf  k} }, {\bf  r})  && \hspace{3cm}
g ^{*} (\omega , - \hat{ {\bf  k} }, {\bf  r}) =
g ( \omega , \hat{ {\bf  k} }, {\bf  r})
\label{eq9}
\end{eqnarray}

In this paper we consider only an order parameter which varies only 
along the $x$ axis. Accordingly $f$ and $g$ depend only on this 
variable. Moreover we assume that the order parameter is periodic in 
this direction which is the situation occuring in the FFLO transition. We 
restrict also ourselves to real order parameters since these have been 
found to correspond to the highest critical temperature in the vicinity of 
the TCP, and the LO solutions are also real, so this property is expected
to be widely satisfied. Anyway the generalization to an intrinsically
complex order parameter should not make much difficulties. 

Then we proceed to a 
Fourier expansion of this order parameter. Let us first assume, in order
to present our method in the simplest case, that only 
the lowest harmonic is relevant. This amounts to take :
\begin{eqnarray}
\Delta (x) = 2 \Delta \cos (qx)
\label{eq10}
\end{eqnarray}
We will consider at the end of the paper the general situation, but we will 
actually find that, for our problem, the actual order parameter at the 
transition is very nearly a 
simple cosine. For fixed $ {\bf  k}$ Eilenberger's equations are a set of 
first order differential equations for the variation of the Green's 
functions along $ {\bf  k}$. 
So we take a reduced variable along this direction by setting 
$ {\bf  r} = {\bf  k} X $, which gives ${\bf  k}.{\bf \nabla} = d/dX$
and $\Delta (x) =  2 \Delta \cos (QX)$ 
where we have introduced $ Q = k _{F} q \cos \theta $ with $ \theta $ 
the angle between $ {\bf  k}$ and the $x$ axis. Then we make a 
Fourier expansion of the Green's functions:
\begin{eqnarray}
f(X) = \sum_{n} f _{n} \: e ^{ inQX}	\hspace{2cm} & f 
^{+}(X) = \sum_{n} f ^{+} _{n} \: e ^{ inQX} & \hspace{2cm}  
g(X) = \sum_{n} g _{n} \: e ^{ inQX}
\label{eq11}
\end{eqnarray}

Explicit substitution of Eq.(11) in Eilenberger's equations Eq.(7) 
gives:
\begin{eqnarray}
f _{n} = \frac{ \Delta }{ \omega + inQ} (g _{n-1}+ g _{n+1}) 
\nonumber  \\
f^{+}_{n} = \frac{ \Delta }{ \omega - inQ} (g _{n-1}+ g _{n+1}) 
\nonumber  \\
g_{n} = \frac{ \Delta }{2inQ} (f _{n-1}+ f _{n+1} - f ^{+}_{n-1}- f 
^{+} _{n+1})
\label{eq12}
\end{eqnarray}

The solution of these equations have simple symmetry properties, 
which can be checked directly for example by generating explicitely 
the solution by a perturbation expansion. 
Actually they arise quite generally from the fact that
we deal with an order parameter which is real and even (i.e.
parity is not broken).
This is more conveniently seen by taking the case where $ \omega  $
is real. However one has to keep in mind that we will deal finally with
a complex $ \omega  $. Nevertheless the symmetry properties are still
valid generally in this case.

For real order parameter $ f(X)$, $ f ^{+}(X)$ and $ g(X)$ are real, 
which is consistent with Eilenberger's equations.
This implies 
$ f _{-n} = f ^{*}_{n}$, $ f ^{+}_{-n} = f^{+*}_{n}$ and $ g _{-
n} = g^{*}_{n}$ . Moreover for an even order parameter, 
Eq.(7) are unchanged when $( \hat{ {\bf  k}},{\bf  
r})$ is changed into $(-\hat{ {\bf  k}},-{\bf  r})$ which shows that $ 
f$, $ f ^{+}$ and $ g$ are also unchanged. Hence from Eq.(9) $ f(-X) 
= f ^{+}(X)$ and $ g(-X) = g(X)$, which leads finally to $ f ^{+}_{n} = f 
_{-n}$ and $ g _{n} = g_{-n}$. 

It is then convenient to make explicit the relation between 
$ f_{n} $ and $ f^{+}_{n}$ by introducing $ d _{n} = (f _{n} -  f^{+}_{n})/2i$, which gives $ f _{n} = (i - 
\omega /nQ ) d _{n}$. We have then for the two quantities $ g _{n}$ 
and $ d _{n}$ (which are real for real $ \omega $) the recursion relations :
\begin{eqnarray}
d _{n} = - \frac{nQ \Delta }{ \omega ^{2}+ n^{ 2} Q ^{2}} (g _{n-1}+ g 
_{n+1})  \nonumber  \\
g _{n} = \frac{ \Delta }{nQ} (d _{n-1}+ d _{n+1})
\label{eq13}
\end{eqnarray}
It is clear from these equations that $ g _{n} \neq 0$ only for even $n$, 
and $ d _{n} \neq 0$ only for odd $n$, as it can be seen for example by generating
the solution perturbatively. Moreover they satisfy   $ g _{-n} =  g _{n}$ and $ d _{-n} =  - d _{n}$.
These equations are linear and 
must be supplemented by the normalization condition Eq.(8). The 
$n=0$ component is enough and it provides us precisely with the 
spatial integral $ g_{0} =  \int \, d{\bf  r} \: g _{s} ( \omega , \hat{ {\bf  
k} }, {\bf  r})$ which we need in Eq.(1) to calculate the free energy :
\begin{eqnarray}
g ^{2}_{0} = 1 - \sum_{n \neq 0} (g_{n} g_{-n} + f _{n} f  ^{+}_{-
n}) = 1 -  \sum_{n = 1} ^{ \infty} (2 g^{2}_{n} +  f ^{2}_{n} +  f ^{2}_{-n})
\label{eq14}
\end{eqnarray}

Now the interesting point is the large $n$ behaviour of $ g _{n}$ and $ 
d _{n}$. If for example we eliminate $ d _{n}$ in Eq.(13) we obtain a 
linear recursion relation which links $ g _{n+2}$ to  $ g _{n}$ and  $ g 
_{n-2}$. Since $ g _{-n} = g_{n}$ we have only to consider $ n \ge 
0$, but this becomes $ n \ge 2 $ when one takes into account that in 
Eq.(13) the relation for $ g _{0}$ is identically satisfied because $ d 
_{-1}= - d _{1}$. The general solution of such a recursion relation is a 
linear combination of two independent solutions. The large $n$ 
behaviour is found from the recursion relation, which for $ \Delta , | 
\omega |  \ll  | nQ | $ simplifies into $ \Delta  ^{2}( g _{n+2} + g _{n-
2}) +  n^{ 2} Q ^{2} g _{n} = 0 $. One sees that this equation has very 
rapidly growing solutions satisfying $ g _{n+2} \gg  g _{n} \gg  g 
_{n-2} $ and behaving as $ g _{2p+2} \sim (-1) ^{p} (2Q/ \Delta 
)^{2p} (p!) ^{2}$. Naturally these solutions are not physically 
acceptable. On the other hand the recursion relation has also a solution 
satisfying $ g _{n+2} \ll  g _{n} \ll  g _{n-2} $ and behaving as $ g 
_{2p} \sim (-1) ^{p} (\Delta / 2Q)^{2p} (1/p!) ^{2}$, which is the 
physical solution we are looking for. This solution is found only if $ g 
_{0}$ and $ g _{2}$ are related by a specific boundary condition. 

The very fast decrease of $ g _{n}$ and $ d _{n}$ provides an easy 
way to obtain a set of approximate solutions, which moreover 
converges rapidly to the exact one, all the more since these are $ g _{n} 
^{2}$ and $ d _{n} ^{2}$ which come in Eq.(14) for the calculation of 
$ g _{0}$. Since  $ g _{n}$ and $ d _{n}$ are very small for large $n$ 
we just take them to be zero beyond some fixed value. This serves as 
boundary condition. Then we work backward to obtain the whole set of 
Fourier components and normalize them properly through the 
normalization condition Eq.(14). Specifically we proceed as follows. 
Since the recursion relations are linear we rescale $ g _{n}$ and $ d 
_{n}$ in order to have convenient initial values. We set $ g _{n} = g 
_{0} G _{n} $ (which implies $G_{ 0}=1 $) and $ d _{n}= nQ  g _{0} D _{n}$ 
and take as initial values $ G _{2N+2} = 0 $ and $ D _{2N+1} 
\neq 0 $ to be determined later. Then starting with $p  = N$ we use for decreasing values of $p$ 
the following recursion obtained from Eq.(13) :
\begin{eqnarray}
G _{2p} = - G _{2p+2} - \frac{ \omega ^{2}+ (2p+1)^{ 2} Q ^{2} }{ 
\Delta } D _{2p+1} \nonumber  \\
D _{2p-1} = \frac{1}{\Delta } [ \frac{2p }{2p-1} G _{2p} - \frac{2p+1 }{2p-1} D 
_{2p+1} ]
\label{eq15}
\end{eqnarray}
down to $ G _{0}$. All the $G$'s and $D$'s are proportional to $ D _{2N+1}$, which is now found by enforcing $G_{ 0}=1 $. Finally Eq.(14) gives explicitely for $ g _{0} $ :
\begin{eqnarray}
{ g ^{-2}_{0}} = 1 + 2 \sum_{p = 1} 
^{N}G^{2}_{2p} + 2 \sum_{p = 0} ^{N} [ \omega ^{2}- 
(2p+1)^{ 2} Q ^{2} ] D^{2}_{2p+1}
\label{eq16}
\end{eqnarray}
When we let $ N \rightarrow \infty $ this equation provides the exact 
result for $g_{0}$. It is interesting to note that for these large $n$ we 
have found that $ g _{n}$  is proportional to $ \Delta ^{n} $. This 
makes a precise link between the expansion in powers of $\Delta$ we 
discussed at the beginning and the Fourier expansion we are 
considering now. One can see our result as corresponding to resummations
of infinite series, eliminating in this way the troubles mentionned in section II occuring
because coefficients in the Landau-Ginzburg expansion change sign as the temperature
is lowered. One finds also that in the limit of large $  | \omega | \gg 
\Delta , | nQ | $, where one must recover the normal state 
Green's functions, one has $ g _{2p} \sim (-1) ^{p} (\Delta / 
\omega)^{2p}$. Naturally the recursion relations Eq.(15) are very 
convenient and very fast for a numerical implementation and in practice 
the situation is not very different from having an analytical expression 
for $ g_{0}$. The only practical problem is linked to the determination 
of the square root in obtaining $ g_{0}$ from Eq.(16), but this is 
solved by noticing that, from the general spectral representation, one 
has $ {\mathrm{Re}} g_{0} \ge 0 $ when  $ \omega _{n} > 0 $.

The simplest of these approximations corresponds to take $N = 0$ and 
it is given explicitely by :
\begin{eqnarray}
g_{0} = [ 1 + 2 \Delta ^{2} \frac{ \omega ^{2} - Q ^{2}}{( \omega 
^{2} + Q ^{2})^{2}} ] ^{- \frac{1}{2} }
\label{eq17}
\end{eqnarray}
This is already a quite non trivial approximation. Since it is correct up to 
order $  \Delta ^{2}$ it gives the proper location for the standard FFLO 
second order transition line. Moreover as we will see it gives 
qualitatively and semiquantitatively the correct results, with a first order 
transition down from the TCP which becomes a second order transition 
at low temperature in agreement with Ref. \cite{matsuo}.

Although it is quite simple the calculation of the free energy has to be 
carried out numerically and naturally it is the same for all the higher 
order approximations. In practice it is convenient to make use in Eq.(1) 
of $  \ln [T/ T _{sp}(\bar{\mu }/T)] = 1/N_{ 0}V - \pi T  \sum {\mathrm sgn( \omega _n)}/ \bar{\omega  
}_{n}$ to rewrite it as : 
\begin{eqnarray}
\frac{ \Omega _{s} - \Omega _{n}}{ N _{0}} =   \ln [T/ T 
_{sp}(\bar{\mu }/T)]  \int d{\bf  r}
 |  \Delta ( {\bf  r}) | ^{2}
+ 4 \pi  T  \sum_{n=0}^{ \infty} \int_{\omega _{n}}^{ \infty}
\, d \omega  \: {\mathrm{Re}} [ < g _{0} ( \omega - i \bar{\mu } , 
\hat{ {\bf  k} }) > _{k}
- 1 + \frac{1}{2 (\omega - i \bar{\mu }) ^{2}}  \int d{\bf  r} |  \Delta ( 
{\bf  r}) | ^{2} ] 
\label{eq18}
\end{eqnarray}
where we have made no assumption on the spatial dependence of $  
\Delta ( {\bf  r}) $. In the present case Eq.(10) gives $  \int d{\bf  r} |  
\Delta ( {\bf  r}) | ^{2} = 2   \Delta ^{2}$. The form Eq.(18) is 
convenient for the frequency integration since the integrant behaves as $ 
\omega ^{-4}$ for large $ \omega $, with a corresponding behaviour $ 
\omega _{n} ^{-3}$ in the Matsubara frequency summation. One may 
replace $  \ln [T/ T _{sp}(\bar{\mu }/T)]$ by $  \ln [\bar{\mu }/ 
\bar{\mu } _{sp}(\bar{\mu }/T)]$ where $  \bar{\mu } _{sp}(\bar{\mu 
}/T) $ is the spinodal field for a given $ \bar{\mu }/T$, since $ 
\bar{\mu } _{sp}(\bar{\mu }/T) / T _{sp}(\bar{\mu }/T) = \bar{\mu 
}/T$. Finally the angular average amounts to an integration over $Q$ 
since $ < g _{0} (\hat{ {\bf  k} }) > _{k} \equiv \int  (d \Omega _{k}/4 
\pi ) g _{0} (\hat{ {\bf  k}}) = \int_{0}^{1}\, du \: g _{0} (Q=\bar{\mu }\bar{q} 
u)$ with $ u = \cos \theta $ and $ \bar{q} = q k _{F}/ \bar{\mu } \equiv 
\hbar q k _{F}/(2m \bar{\mu })$.

Since we are only interested in the transition from the normal to the 
FFLO state, in order to obtain the corresponding critical temperature $ T 
_{c}$ we look for the highest temperature $T$ (or effective field $ 
\bar{ \mu }$) at which $  \Omega _{s} - \Omega _{n} = 0 $. Precisely 
this leads to the following equation for $ T = T _{c}$:
\begin{eqnarray}
\ln [T / T _{sp}(\bar{\mu }/T)] = - {\mathrm{Min}} \frac{2 \pi  T 
}{\Delta ^{2}} 
\sum_{n=0}^{ \infty} \int_{\omega _{n}}^{ \infty}
d \omega \, {\mathrm{Re}} [ < g _{0} ( \omega - i \bar{\mu } , \hat{ 
{\bf  k} }) > _{k}
- 1 + \frac{ \Delta ^{2}}{(\omega - i \bar{\mu }) ^{2}}] 
\label{eq19}
\end{eqnarray}
Since for homogeneity the right-hand side of this equation is only a 
function of $ T / \bar{\mu }$, $ \Delta / \bar{\mu }$ and $ \bar{q} / 
\bar{\mu } $ one has to minimize with respect to $ \Delta / \bar{\mu }$ 
and $ \bar{q} / \bar{\mu } $, at fixed $ T / \bar{\mu }$.

At $ T = 0$ this simplifies somewhat. The summation over Matsubara 
frequencies goes to an integral which is performed by a by parts 
integration. This gives for the critical field $ \bar{\mu }$ :
\begin{eqnarray}
\ln [\bar{\mu }/ \bar{\mu } _{sp}(T=0)]  = - {\mathrm{Min}} \frac{1 
}{\Delta ^{2}} 
\int_{0}^{\infty }
d \omega \, \omega {\mathrm{Re}} [ < g _{0} ( \omega - i \bar{\mu } , 
\hat{ {\bf  k} }) > _{k}
- 1 + \frac{ \Delta ^{2}}{(\omega - i \bar{\mu }) ^{2}}]
\label{eq20}
\end{eqnarray}

Finally we write down the self-consistency equation (or 'gap equation') for the order parameter, which comes out quite generally when the free energy is minimized with respect to variations of this order parameter. This equation gives the necessary feedback to Eq.(\ref{eq7}) where $\Delta ( {\bf  r})$ can not be naturally an open function. In practice we will make little use of it since we will rather minimize directly the free energy with respect to the order parameter. The self-consistency equation \cite{eil1}  can be written as:
\begin{eqnarray}
\Delta _{n }\ln [T / T _{sp}(\bar{\mu }/T)] = 2 \pi  T 
\sum_{m=0}^{ \infty}  \, {\mathrm{Re}} [ < f _{n} ( \omega - i \bar{\mu } , \hat{ 
{\bf  k} }) > _{k}
- \frac{\Delta _{n }}{\omega - i \bar{\mu }}] 
\label{eqgap}
\end{eqnarray}
by making the same transformation from $T_{c0}$ to $ T _{sp}(\bar{\mu }/T)$ as we did to find Eq. (\ref{eq18}).

\section{GENERAL ORDER PARAMETER}

We consider now the extension of our Fourier expansion, presented in section \ref{fourier}, to a general order parameter:
\begin{eqnarray}
\Delta (x) =  \sum_{n}  \Delta_{n} \: e ^{ inQX}
\label{dgen}
\end{eqnarray}
We assume again a real order parameter, which implies $\Delta^{*} _{n} = \Delta_{-n}$. We also restrict ourselves as before to an order parameter even with respect to $x$, which makes $\Delta_{n}$ real.
Substitution as above in Eilenberger's equations give the following generalization of Eq.(\ref{eq12}):
\begin{eqnarray}
(\omega + inQ)f _{n} =  (\omega - inQ)f^{+}_{n} = \sum_{p=1}^{\infty} \Delta_{ p} (g _{n-p}+ g _{n+p}) 
\nonumber  \\
2inQ g_{n} =  \sum_{p=1}^{\infty} \Delta_{ p} (f _{n-p}+ f _{n+p} - f ^{+}_{n-p}- f ^{+} _{n+p})
\label{fourgen}
\end{eqnarray}
Here we have also assumed for the moment that the spatial average of the order parameter is zero, that is $ \Delta_{ 0}=0$. Introducing again $ d _{n} = (f _{n} -  f^{+}_{n})/2i$, we obtain the recursion relations:
\begin{eqnarray}
d _{n} = - \frac{nQ}{ \omega ^{2}+ n^{ 2} Q ^{2}} \sum_{p=1}^{\infty} \Delta_{ p} (g _{n-p}+ g _{n+p})   \nonumber  \\
g _{n} = \frac{1 }{nQ}  \sum_{p=1}^{\infty} \Delta_{ p} (d _{n-p}+ d _{n+p})
\label{recurgen}
\end{eqnarray}

We consider now the practical situation met in numerical use of these equations. In this case the number of Fourier components for the order parameter will be finite, so we have a maximum integer $P$ for which  $\Delta_{ p} = 0$ when $ p > P$. We look again at the behaviour of the physical solution for $ d_{ n}$ and $g_{ n} $ when $n$ goes to infinity, and show that it is consistent with a fast factorial type decrease, as we have found in section \ref{fourier}. Indeed in this case the dominant term in the sums found in the right hand side of Eq.(\ref{recurgen}) is the one where  $ d_{ n}$ or $g_{ n} $ has the smallest index $n$, which gives $ d_{ n} \sim - \Delta_{ P} g_{ n-P}/(nQ)  $ and $g_{ n} \sim  \Delta_{ P} d_{ n-P}/(nQ)$. This leads to $g_{ n} \sim ( \Delta_{ P}/Q)^{ n/P}/(n!)^{ 1/P}$. Although this is still a fast factorial type decrease, it gets slower when $P$ increases. On the other hand the large $n$ behaviour contains the power law dependence $ ( \Delta_{ P})^{ n/P}$, where in generic situations $ \Delta_{ P} $ is expected to be very small for large $P$. This is indeed what is found when one writes the self-consistency equation \cite{eil1} Eq.(\ref{eqgap}) which gives  $ \Delta_{ n} $ in terms of $ f_{ n} $. This fast decrease of $ \Delta_{ n} $ corresponds to the standard situation, where there is no smaller physical length scale for $ \Delta (x)$ than $1/q$ itself. However one can think of other particular situations, where this fast decrease of $ \Delta_{ n} $ does not hold and which should be dealt with specifically. Ultimately this convergence problem has to be handled numerically by making calculations for increasing $P$ and looking when reasonable convergence has been achieved. This is what we will do below with our present problem of finding the location of the transition.

Finally we make the same practical use of this fast convergence property as in section \ref{fourier}. We take as boundary condition that $g_{ n}$ and $ d_{ n}$  are zero beyond some fixed value $N$. This allows to calculate all the $g_{ n}$ and $ d_{ n}$ within a common multiplicative factor, which is then found by the normalization condition Eq.(\ref{eq14}). $N$ is progressively increased until convergence has been obtained. The situation for solving the practical problem of finding the $g_{ n}$'s and the $ d_{ n}$'s is less convenient than in section \ref{fourier}. However we still have a linear problem for which very efficient numerical procedures are known. We have basically to handle a matrix. Instead of having a tridiagonal matrix, with just matrix elements right below and above the main diagonal we have now a band diagonal matrix with, in addition to the main diagonal, $2P$  diagonals with non zero matrix elements.

To be more specific we have now to take into account that, in our problem, only odd Fourier components $\Delta _{ 2p+1} $ of the order parameter are nonzero. First we consider only order parameters with a zero spatial average $ \Delta_{ 0}  =0$, since taking a nonzero value amounts to mix in the order parameter of the uniform BCS phase, which is energetically unfavorable. Hence it is reasonable to assume that similarly a nonzero $ \Delta_{ 0} $ is unfavorable. Next we can see for example by a iterative treatment to all orders, in order to obtain an exact solution of Eilenberger's equations, that we have only odd components. Indeed if we start with the simple $\Delta (x) = 2 \Delta \cos (qx)$ that we considered in section \ref{fourier}, we generate only odd Fourier components in $f(X)$ and even components in $g(X)$ as we have seen. Now this $f(X)$ in turn generates only odd components for $\Delta (x)$ from the self-consistency equation Eq.(\ref{eqgap}). But from Eq.(\ref{recurgen}) this is again completely compatible with only odd components for $f(X)$ and even for $g(X)$. Naturally it can also be seen directly from the starting Eilenberger's equations that such a solution is a consistent one. We note that such a solution with odd components means that, by shifting $x$ by $ \pi /(2q)$, we obtain an order parameter which is odd with respect to $x$, in the same way as it transforms Eq.(\ref{eq10}) into $2 \Delta \sin (qx)$.

Then it results from Eq.(\ref{recurgen}) that, just as in section \ref{fourier}, $ g _{n} \neq 0$ only for even $n$, and $ d _{n} \neq 0$ only for odd $n$. In the same way we set $ g _{n} = g 
_{0} G _{n} $ (implying $G_{ 0}=1 $) and $ d _{n}= nQ  g _{0} D _{n}$, and we have again $ G _{-n}= G _{n}$ and $ D _{-n}= D _{n}$. It is now convenient to include $ g_{ n} $ and $d_{ n} $ into a single unknown vector $V_{ n} $, defined by $V_{2p}= G_{ 2p}  $ and $V_{2p+1}= D_{ 2p+1}  $. Then Eq.(\ref{recurgen}) can be merely written as $ M_{mn} V_{n} = A_{m}   $ with $A_{n}=- \Delta _{n}$ and the matrix $M$ given by:
\begin{eqnarray}
M_{2n,2n} = 2n \nonumber \\
M_{2n+1,2n+1} = \omega ^{2} + (2n+1)^{ 2} Q ^{2} & & \nonumber  \\
M_{2m+1,2n} = \Delta _{2(m+n)+1} + \Delta _{|2(m-n)+1|} & & \nonumber  \\
M_{2m,2n+1} = (2n+1) [ \Delta _{2(m+n)+1} - \Delta _{|2(n-m)+1|}] & &
\label{}
\end{eqnarray}
with $m,n \ge 1$. 

As explained above we truncate the infinite matrix $M$ by $m,n \le N_{\rm max}$, which gives a matrix with finite order $N_{\rm max}$. The corresponding linear equation for $V_{n}$, with $n \le N_{\rm max}$, can be solved numerically by efficient standard routines, since as mentionned above the matrix $M$ has a generalized band diagonal form. Once this is done, $g_{0}$ is still obtained from Eq.(\ref{eq16}), the free energy calculated from Eq.(\ref{eq18}) and the critical temperature obtained by minimization. Finally the whole procedure is repeated for increasing values of $N_{\rm max}$ until convergence is achieved. In the next section we will display the corresponding numerical results.

\section{NUMERICAL RESULTS}\label{num}

We present now the results of our numerical implementation of the above procedure. In the first susbsection below we restrict ourselves to an order parameter with only the lowest harmonic as it is given by Eq.(\ref{eq10}). The general case is considered afterwards.

\subsection{Lowest harmonic}

We first give in Fig. 1 the results for the calculation of the critical temperature, down from the TCP. Rather than giving $T_c ( \bar{\mu }) $, we cover for convenience the ($ \bar{\mu },T$) plane in polar coordinates, rather than cartesian coordinates, and give the critical temperature $T_c ( \bar{\mu } / T) $ as a function of $ \bar{\mu }/T$, equivalent to a polar angle. More precisely we plot its ratio $T_c/T_{ FFLO}$ to the FFLO critical temperature $T_{ FFLO}( \bar{\mu } / T)$ obtained for the same value of the ratio $  \bar{\mu } / T$. This is more suited to the present situation since we find this ratio to be always near unity. However to make the graph more readable, we give on the x-axis,  instead of  $  \bar{\mu } / T$, the value of $T_{ FFLO}( \bar{\mu } / T)$ itself, compared to the standard BCS critical temperature $T_{ c0} $, found for $ \bar{\mu}=0$ (this corresponds to go along the standard FFLO transition line). Naturally when our result for $T_c/T_{ FFLO}$ goes below $1$, this means that the first order transition is less favorable than the second order one, so when the temperature is lowered there is actually a switch from first to second order when one finds that $T_c/T_{ FFLO}=1$. 

We give the results of the calculation with increasing values of $N_{\rm max}$ going up to 6. The approximation $N_{\rm max}=1$ corresponds to the explicit result Eq.(\ref{eq17}) for $g_0$. As already mentionned above, it is correct up to second order in $\Delta $ and consequently it gives the correct location for the FFLO transition. Moreover we see that it gives already the proper result semi-quantitatively for the order of the transition, since it is gives a switch from first to second order when the temperature goes below $ T_{ FFLO} / T_{ c0} = 0.195 $. The next approximation $N_{\rm max}=3$ for odd $N_{\rm max}$ is already quite good quantitatively since it gives $0.063$ for the above ratio. Full convergence is obtained for $N_{\rm max}=5$ where we find $ T_{ FFLO} / T_{ c0} = 0.076 $ in very good agreement with Matsuo \emph{et al} \cite{matsuo}. For completeness we give also in Fig. 1 our results for even $N_{\rm max}$. It is less natural, from the structure of the recursion equations, to truncate them in this way. Hence it is not so surprising that the approximation $N_{\rm max}=2$ is much worse than $N_{\rm max}=1$ since it does not even give a switch from first to second order for the transition. Nevertheless we have naturally convergence when we increase $N_{\rm max}$, and indeed we find that $N_{\rm max}=4$ is already very good since the switch is located at $ T_{ FFLO} / T_{ c0} = 0.074 $, while $N_{\rm max}=6$ is completely converged.
\begin{figure}[htbp]
\begin{center}
\vbox to 60mm{\hspace{-6mm} \epsfysize=60mm \epsfbox{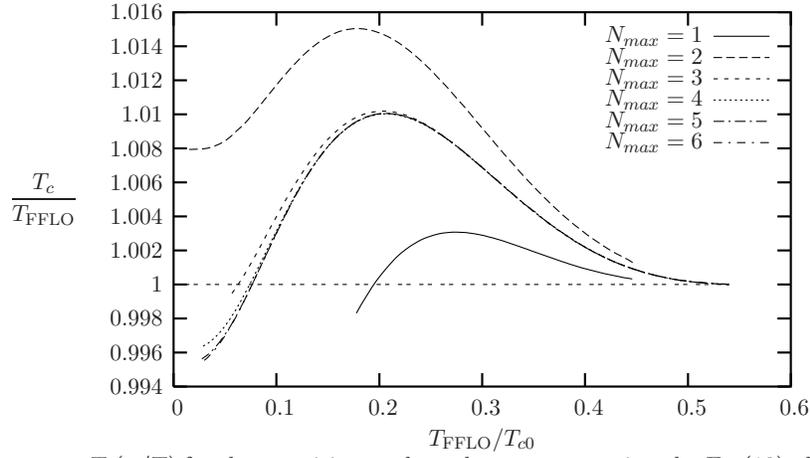}}
\caption{Critical temperature $T_c ( \bar{\mu } / T) $ for the transition to the order parameter given by Eq.(10), divided by the FFLO critical temperature $T_{ FFLO}( \bar{\mu } / T)$ obtained for the same value of  $  \bar{\mu } / T$. On the $x$ axis, instead of $  \bar{\mu } / T$, we have given $T_{ FFLO}( \bar{\mu } )/ T_{ c0} $, where $T_{ c0} $ is the maximal critical temperature, obtained for $ \bar{\mu}=0$.}
\label{default}
\end{center}
\end{figure}
A noticeable feature of Fig. 1 is that the ratio $T_c/T_{ FFLO}$ stays always very near unity, while one would have expected to find it larger since there is no obvious relation between the order parameters of the first and the second order transitions. This behaviour is also found \cite{cm} near the TCP. A natural conclusion from this feature is to say that the first order transition is actually always quite near to be a second order one. We can check in our results if this interpretation is a coherent one by looking at the size of the order parameter (more precisely its maximal value as a function of spatial position), that is essentially the value of $ \Delta _{1}$ at the first order transition. If the first order transition is nearly a second order one, it should be small compared to a typical gap $ \Delta _{0}$ in the uniform BCS phase. Equivalently, in Fig. 2, we compare it to $\bar{\mu }$ since it is of order $ \Delta _{0}$ in all the range we are interested in (at $T=0$ the FFLO result is $\bar{\mu } = 0.754 \Delta _{0}$). Our results show clearly that the size of the order parameter at the transition is of order of the one deep in the standard BCS phase, so it is not at all possible to consider that the first order transition is nearly a second order one.
\begin{figure}[htbp]
\begin{center}
\vbox to 60mm{\hspace{-6mm} \epsfysize=60mm \epsfbox{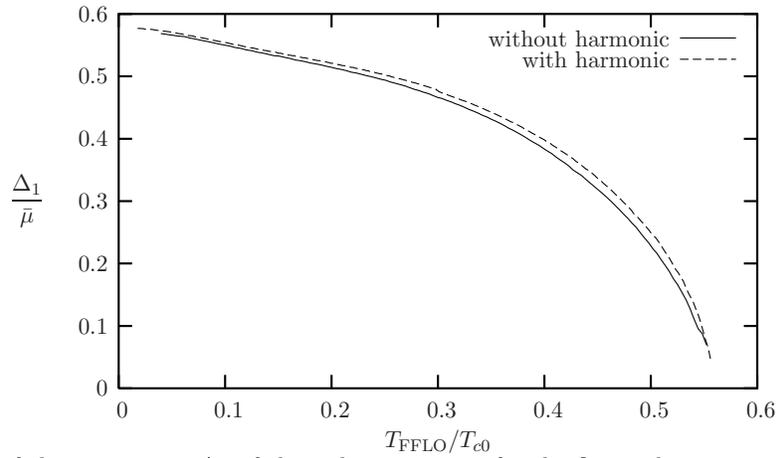}}
\caption{Full line: size of the component $ \Delta _{1}$ of the order parameter for the first order transition, compared to $\bar{\mu } $, when all higher order harmonics are taken equal to zero. Dashed line: same quantity for an order parameter where $ \Delta _{3}$ is also non zero.}
\label{default}
\end{center}
\end{figure}
Finally it is also of interest to compare the reduced wavevectors of the order parameter for our first order transition and for the standard second order FFLO transition. This is done in Fig. 3 where it is seen that, although there are differences, they not very large so that the reduced wavevector is rather similar for the two transitions, in contrast with the size of the order parameter.
\begin{figure}[htbp]
\begin{center}
\vbox to 60mm{\hspace{-6mm} \epsfysize=60mm \epsfbox{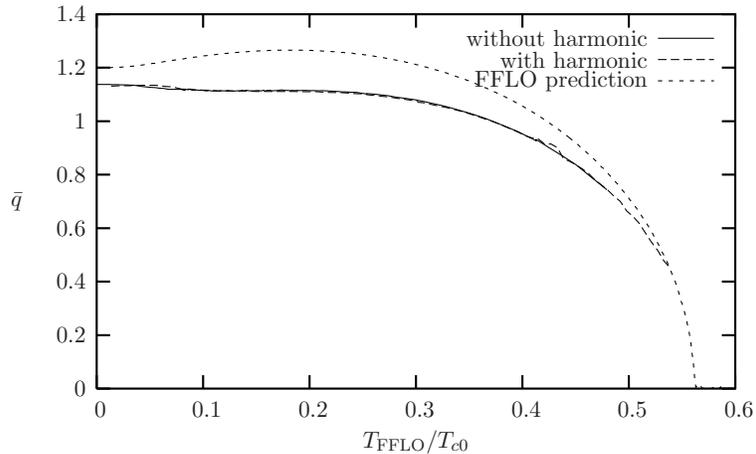}}
\caption{Full line: optimal reduced wavevector $\bar{q}$ of the order parameter at the first order transition, for the converged solution, as a function of temperature, when only the component $ \Delta _{1}$ is different from zero. Long dashed line: same result  when $ \Delta _{3}$ is also non zero. Short dashed line: corresponding result for the second order FFLO transition}
\label{default}
\end{center}
\end{figure}

\subsection{Higher harmonics}

Naturally it is not consistent to keep only the lowest harmonic in the order parameter, as it is immediately seen from the self-consistency equation (\ref{eqgap}). Hence we consider now the effect of higher harmonics. In a first step we have explored the effect at the transition of the inclusion of  $ \Delta _{3}$. We have found it quite small. This would imply normally to stop the exploration at this stage since one expects the effect of harmonics $ \Delta _{5}$ and higher to be even smaller. However one might wonder whether this result is not somewhat accidental \cite{buzcom} and specific to $ \Delta _{3}$. This is especially a concern in the vicinity of the switching temperature from first to second order, where $ \Delta _{3}$ is particularly small (see Fig. \ref{delta5}). It could be that higher order harmonics dominate in this region
leading to a quantitative change of the first order transition line. Hence, in order to eliminate any doubt on this question, we have explored the effect of taking $ \Delta _{5}$ and $ \Delta _{7}$ to be different from zero. Our study shows that these harmonics give also a very small contribution. Hence our conclusion is that the optimal order parameter remains always very close to a simple cosine in the whole temperature range from the tricritical point down to zero temperature.

Our numerical procedure is to use directly the free energy Eq.(\ref{eq18}) by taking as an ansatz our form for the order parameter, with either three or four odd Fourier components. More precisely we maximize, with respect to $ \bar{q}, \Delta _{\rm i}$ with ${\rm i}=1,3,5,7$, the critical temperature from the generalization to our case of Eq.(\ref{eq19}), as explained in section \ref{fourier}. We then check that our optimal form satisfies the gap equation. We have also performed calculations by making use only of the gap equation. The results are not significantly different from the ones we display below, and most of the time agree with them within numerical accuracy . From a practical point of view, we have chosen high enough values of $N_{\rm max}$, typically  $N_{\rm max} = 12$, so that numerical results are insensitive to changes in  $N_{\rm max}$.

\begin{figure}[htbp]
\begin{center}
\vbox to 60mm{\hspace{-6mm} \epsfysize=60mm \epsfbox{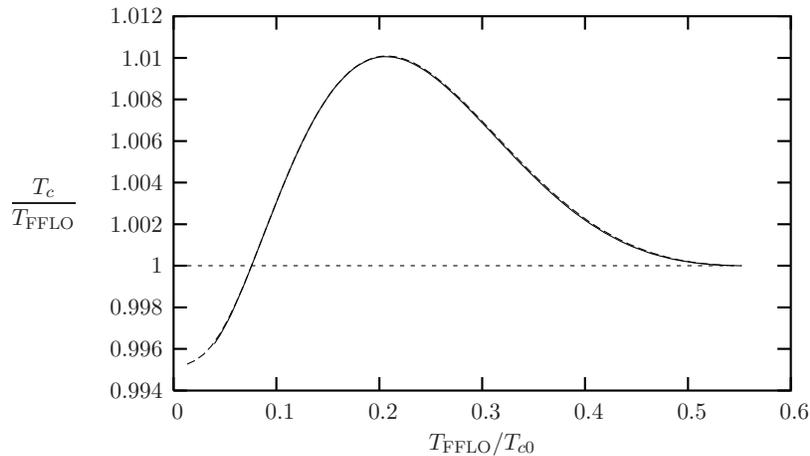}}
\caption{Dashed line: critical temperature for the first order transition for a one dimensional order parameter form with four odd Fourier component $\Delta _{\rm i}$ with ${\rm i}=1,3,5$ and $7$. Full line: same result for the simple cosine ansatz, where only  $\Delta _{1}$ is different from zero.
\label{figTcrit2}}
\end{center}
\end{figure}

\begin{figure}[htbp]
\begin{center}
\vbox to 60mm{\hspace{-6mm} \epsfysize=60mm \epsfbox{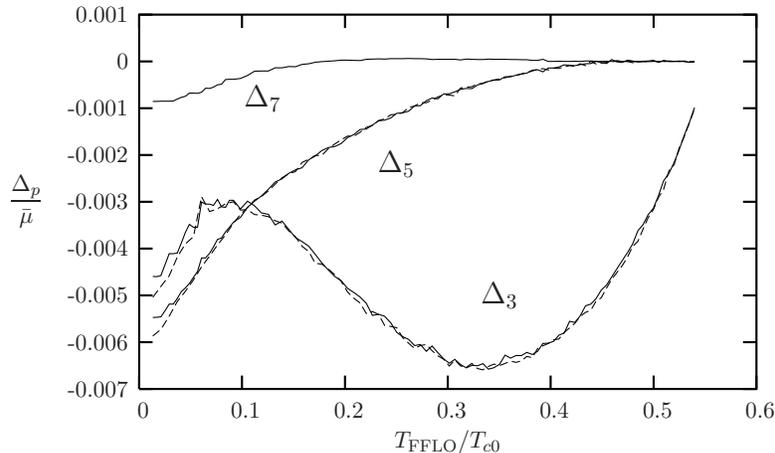}}
\caption{Optimal values for the amplitudes $\Delta_3$,
 $\Delta_5$  and $\Delta_7$ in the order parameter as a function of temperature
(full lines). For comparison the results for $\Delta_3$ and  $\Delta_5$ when $\Delta_7 = 0$
are also given as dashed lines.
\label{delta5}}
\end{center}
\end{figure}

We give first in Fig. \ref{figTcrit2} our result for the critical temperature of the first order transition. The effect of all our higher order harmonics can be only barely seen in the figure, as compared to our calculation with only the lowest harmonics $\Delta _{1}$, already given in Fig. 1.
Next we display in Fig. \ref{delta5}, as a function of temperature, the values of the higher order harmonics $\Delta _{\rm 3}$,$ \Delta _{\rm 5}$ and $\Delta _{\rm 7}$ for the optimal order parameter. It is seen that they are always quite small compared to $\Delta _{\rm 1}$. Nevertheless, around and below the switching temperature, $\Delta _{\rm 3}$ and $ \Delta _{\rm 5}$ are of the same order while one would have rather expected $\Delta _{\rm 5}$ to be small compared to $ \Delta _{\rm 3}$ (note that anyway these results are physically irrelevant below the switching temperature since they are for the first order transition, while the actual transition is second order). On the other hand $\Delta _{\rm 7}$ is always negligible compared to $\Delta _{\rm 3}$ and $\Delta _{\rm 5}$, except near the TCP where anyway $\Delta _{\rm 5}$ and $\Delta _{\rm 7}$ are essentially zero. Finally  Fig. \ref{delta5} shows also the results for the optimal $\Delta _{\rm 3}$ and $ \Delta _{\rm 5}$ when $\Delta _{\rm 7} = 0$. The difference with the preceding results is within numerical error. Similarly our result for $\Delta _{\rm 3}$ when $\Delta _{\rm 5} = \Delta _{\rm 7} = 0$ (not shown) are also essentially indistinguishable from the result displayed in Fig. \ref{delta5}.

\section{ CONCLUSION }

In this paper we have shown that performing a Fourier expansion in the quasiclassical Eilenberger's equations provides a very efficient way to study the transition from the normal state to the FFLO phases in 3 dimensions, at least in the vicinity of the transition. We have applied this technique to the case of the transition to the one-dimensional 'planar' order parameter and we have found perfect agreement with the earlier work of Matsuo \emph{et al}. In particular we have rederived their remarkable result that, when the temperature is lowered, the transition switches from first to second order. We have shown in detail that, in the case of the first order transition, the order parameter is nevertheless dominated by its lowest order Fourier component, in somewhat surprising contrast to what one might guess for such a transition. This feature contributes naturally to make our Fourier expansion very rapidly converging.

However the strength of our method is not so much displayed in this case of a one-dimensional order parameter. Its major interest is rather that our approach can be fairly easily generalized to more complex order parameters, with full three-dimensional spatial dependence. As shown by Larkin and Ovchinnikov these order parameters come in competition and, in the case of a first order transition, it is not clear that they are not more advantageous than the standard second order FFLO phase transition. We will indeed show, in forthcoming work, that this is the case at low temperature in 3 dimensions. Finally another interest of our approach is to provide some insight, even if approximate, in the analytical structure of the theory, as we have seen by providing explicit approximate analytical solutions. We believe that this might be helpful in a theoretical situation where the intrinsic non linearity of the equations forces mostly to purely numerical work. 

\vspace{4mm} 
* Laboratoire associ\'e au Centre National
de la Recherche Scientifique et aux Universit\'es Paris 6 et Paris 7.

\end{document}